\documentclass{elsart3p}

\usepackage{graphics}
\usepackage{epsfig}

\usepackage{amssymb}

\begin{document}

\begin{frontmatter}

% Title, authors and addresses

% use the thanksref command within \title, \author or \address for footnotes;
% use the corauthref command within \author for corresponding author footnotes;
% use the ead command for the email address,
% and the form \ead[url] for the home page:
% \title{Title\thanksref{label1}}
% \thanks[label1]{}
% \author{Name\corauthref{cor1}\thanksref{label2}}
% \ead{email address}
% \ead[url]{home page}
% \thanks[label2]{}
% \corauth[cor1]{}
% \address{Address\thanksref{label3}}
% \thanks[label3]{}

\title{Sign reversal of the Hall resistance in the mixed-state of
La$_{1.89}$ Ce$_{0.11}$CuO$_{4}$ and
La$_{1.89}$Ce$_{0.11}$(Cu$_{0.99}$Co$_{0.01}$)O$_{4} $ thin films}

% use optional labels to link authors explicitly to addresses:
% \author[label1,label2]{}
% \address[label1]{}
% \address[label2]{}

\author[label1]{K. Jin},
\author[label1]{B. X. Wu},
\author[label1]{B. Y. Zhu}$^{\star}$,
\author[label1]{B. R. Zhao},
\author[label2]{A. Volodin},
\author[label2]{J. Vanacken},
\author[label2]{A. V. Silhanek},
\author[label2]{V. V. Moshchalkov}

\address[label1]{National Laboratory for Superconductivity, Institute of
Physics, and Beijing National Laboratory for Condensed Matter
Physics, Chinese Academy of Sciences, Beijing 100080, China}

\address[label2]{INPAC - Institute for Nanoscale Physics and Chemistry,
K. U. Leuven, Celestijnenlaan 200D, B-3001 Leuven, Belgium}
\date{\today }

\begin{abstract}
The transport properties of La$_{1.89}$Ce$_{0.11}$CuO$_{4}$(LCCO)
and La$_{1.89}$Ce$_{0.11}$(Cu$_{0.99}$Co$_{0.01}$)O$_{4}$
(LCCO:Co) superconducting thin films are investigated. When the
external field $\bf H$ is applied along the crystallographic
$c$-axis, a double sign reversal of the Hall voltage in the mixed
state of LCCO:Co thin films is observed whereas a single sign
reversal is detected in LCCO. A double sign reversal of the Hall
signal in LCCO can be recovered if the magnetic field is tilted
away from the plane of the film. We find that the transition from
one to two of the Hall sign reversal coincides with the change in
the pinning from strong to weak. This temperature/field induced
transition is caused either by the magnetic impurities in LCCO:Co
or by the coupling between the pancake vortices and the in-plane
Josephson vortices in LCCO. These results are in agreement with
early theoretical and numerical predictions.
\end{abstract}

\begin{keyword} Flux pinning \sep magnetic properties \sep cuprate superconductors \sep superconducting films
% keywords here, in the form: keyword \sep keyword

% PACS codes here, in the form: \PACS code \sep code
\PACS 74.25.Qt \sep 74.25.Ha \sep 74.72.-h \sep 74.78.-w
\end{keyword}
\end{frontmatter}

%, 78.20.Bh
%74.25.Qt, Vortex lattices, flux pinning, flux creep
%74.25.Ha, Magnetic properties
%74.72.-h, Cuprate superconductors (high-Tc and insulating parent compounds)
%74.78.-w, Superconducting films and low dimensional structures
%78.20.Bh, Theory, models, and numerical simulation

% main text
%\label{}

\section {Introduction}

Hall effect is considered as a powerful method to probe the Fermi
surface of metallic compounds in general, and particularly useful to
identify the nature of the charge carriers in non-magnetic
systems\cite{Oh-prl-2007, singleton-book, balakirev, leboeuf,
pfleid}. However, in the mixed state of superconductors, the Hall
voltage is mainly determined by the vortex motion along the
direction of the bias current flow\cite{vgfb}. In these systems, a
long standing unsolved issue is the presence of a sign reversal in
the Hall voltage when changing either temperature or magnetic field.

There have been many experiments and theoretical studies focused
on this so called anomalous Hall effect. For example, sign
reversals has been reported in YBa$_{2}$Cu$_{3}$O$_{7}$ (YBCO)
crystals~\cite{kang96, gob} and La$_{2-x}$Sr$_{x}$CuO$_{4}$ thin
films~\cite{matsuda95}, and double sign reversal and even triple
sign reversal have been observed in
Tl$_{2}$Ba$_{2}$CaCu$_{2}$O$_{8}$ (TBCCO)~\cite{hagen91} and
HgBa$_{2}$CaCu$_{2}$O$_{6}$ thin films~\cite{kang}, respectively.
Several models, based on two-band~\cite{hirsch91}, thermal
fluctuation~\cite{aronov90}, pinning effect~\cite{vgfb, wdt}, and
vortex interaction~\cite{zhu,roger}, have been proposed to
interpret the Hall anomalies. In electron-doped High-T$_c$
superconductors, a distinct feature revealed by the
angular-resolution photoemission experiments~\cite{armit} is the
coexistence of the hole- and electron-bands near the optimal
doping region, which may account for the sign reversal of the Hall
resistivity for temperatures above the superconducting transition
$T_c$~\cite{kui07B}. However, for explaining the presence of sign
reversals below $T_c$, it is still necessary to discuss them in
terms of vortex pinning mechanisms which have been demonstrated to
play a key role~\cite{wdt, zhu}.

Indeed, in a seminal work, Wang and Ting~\cite{wt91} demonstrated
that the Hall resistivity $\rho_{xy}$ as a function of the
magnetic field in the flux flow regime exhibits a sign change due
to backflow currents arising from pinning forces. However,
Vinokur, Geshkenbein, Feigel'man, and Blatter~\cite{vgfb} on the
basis of a phenomenological model which included pinning and
thermal fluctuation, concluded that the magnitude of the Hall
angle $\Theta_H$, does not depend on the pinning. Whether pinning
is necessary for the anomalous Hall effect or not is still a
fundamental question, although there is clear experimental
evidence indicating that the abnormal Hall effect is indeed
pinning dependent~\cite{kang96}.

Based on their former work, Wang, Dong, and Ting (WDT) developed a
unified theory for the flux motion~\cite{wdt} by taking into
account vortex pinning and thermal fluctuations which could
explain both the single and the double sign reversals of the Hall
resistivity due to different pinning strength. The WDT prediction
has been further confirmed by Zhu {\it et al.} by a numerical
simulation~\cite{zhu96}. Their work has clearly shown that the
double sign change of $\rho_{xy}$ versus the temperature appears
in the weak pinning environment, while the single sign reversal
occurs in the strong pinning system. Importantly, the authors have
also found that there are two distinctive weak pinning
configurations which can achieve the double sign reversal of
$\rho_{xy}$(T), i.e., one possesses low density of the pinning
centers but with strong individual pinning potentials, whereas the
other consists of a high pinning concentration of weak individual
pinning potentials. An elegant experimental study by Kang {\it et
al.}~\cite{kang96} unambiguously showed the relevance of the
pinning strength for the sign reversal of the Hall resistance by
irradiating a sample with heavy-ion to progressively increase the
pinning strength. What it is still lacking is a more tunable
pinning which allows one to go from the strong limit to the weak
limit. In this case, one may expect a transition from the single
sign reversal to the double sign reversal in the Hall resistivity
$\rho_{xy}$.

In the present work, we present experimental evidence on the
relevance of the pinning strength in determining the number of
sign reversals in the Hall voltage. We show that despite the fact
that LCCO thin films exhibit only one sign reversal with the
applied field perpendicular to the $ab$-plane, a second sign
reversal emerges either when rotating the magnetic field and
maintaining the same perpendicular component or by changing the
pinning strength by substituting Co for Cu in order to obtain
La$_{1.89}$Ce$_{0.11}$(Cu$_{0.99}$Co$_{0.01}$)O$_{4}$ (LCCO:Co).
These two different methods show that the transition of the Hall
sign reversals from single to double, coincides with the change in
the pinning environment from strong to weak. Our experimental
results are in agreement with previous prediction based on
analytical and numerical investigations~\cite{wdt, zhu96}.

\section {Experimental details}

All the LCCO thin films used in the present work were fabricated
by a dc magnetron sputtering method~\cite{wbx1, wbx2}. The
transport measurements of both LCCO and LCCO:Co thin films were
carried out by using a commercial Quantum Design PPMS-14. More
details about the film preparation and the measurement setup can
be found elsewhere~\cite{kui07B, wbx1, wbx2, kui06B, kui07C, kui08B, kui11}.
All the thin films were patterned by photolithography and subsequent ion
milling, into bridges of $2100~\mu m$ long, $100~\mu m$ wide and
with six terminals. The critical transition temperatures $T_{c0}$
for LCCO and LCCO:Co are $\sim 24$ and 13~K as reported in 
our previous work~\cite{kui06B}, respectively. The upper critical 
magnetic field $H_{c2}(0)$ for LCCO is about 10~Tesla~\cite{kui07B}.

\section {Results and discussions}

% start LCCO
Let us start by demonstrating experimentally that the tilting of
the magnetic field with respect to the plane of a thin LCCO film
leads to significant changes in the pinning properties as well as
the sign reversal of the Hall resistivity $\rho_{xy}$. In order to
investigate this effect we study the Hall resistivity $\rho _{xy}$
and longitudinal resistivity $\rho_{xx}$ as a function of the
perpendicular field $B_z$ by either varying the strength but
maintaining the field orientation perpendicular to the $ab$-plane
(mode A) or by changing the angle $\theta$ between the field and
the $ab$-plane with constant field intensity (mode B). During the
measurements, the applied current $j$ in the $ab$-plane is always
perpendicular to the magnetic field, i.e., $\bf{j} \perp \bf{B}$.
This method allows us to discern the effects arising purely from
the out of the plane component of the applied field from those
originated by the in-plane field.

\begin {figure}[!tbp] %fig 1
\begin{center}
\includegraphics*[bb=35 353 443 746, width=7.5cm, clip]{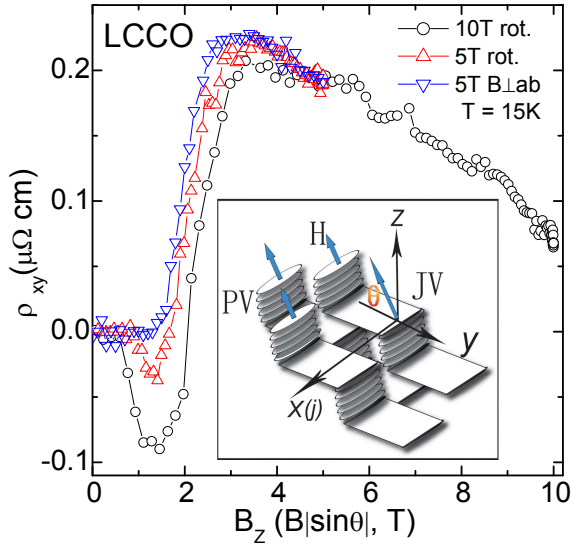}
\end{center}
\caption{(Color online) Hall resistivity $\rho_{xy}$ versus $B_z$
($B|\sin \theta|$) in LCCO thin films at 15~K by different field
changing modes. Inset: illustration of tilted vortices with JV
representing in-plane Josephson vortices and PV representing the
pancake vortices. } \label{Bz}
\end {figure}

First we would like to point out that in the LCCO sample and for
$T>T_{c}$, both $\rho _{xy}(B_z)$ and $\rho_{xx}(B_z)$ curves
overlap with each other no matter whether the perpendicular
component of the magnetic field changes its magnitude or its
orientation~\cite{kui07B}. This indicates that the contribution of
parallel-field component ($B_{y}$) to the magnetoresistance is
negligible in comparison with $B_z$. The question naturally arises
whether this is also true for the case of $T<T_{c0} $. In
Fig.~\ref{Bz}, isothermal curves of $\rho _{xy}$ versus
$B_z=B|\sin \theta |$ for LCCO are plotted at 15~K ($<T_{c0})$.
Here, we find that $\rho _{xy}(B_z)$ curve does not follow the scaling 
with $B_z$ as well as the $\rho _{xx}(B_z)$ curve as reported 
in Ref.~\cite{kui07B}. 
Notice that $\rho
_{xy}(B_z)$ with $\bf{B}\perp ab$-plane has the maximum value at
$\sim 3.5$~T and drops to zero when the magnetic field is
decreased down to 1.4~T without reversing its sign, whereas the
$\rho _{xy}(B)$ measurement via rotation of the field exhibits a
clear sign change from positive to negative in the low field
region.

%--------------------------------------------------------------------
%--------------------------------------------------------------------

This result indicates the importance of the in-plane component of
the field in the mixed state. When the magnetic field is titled
away from the $c$-axis, two different kinds of vortices appear in
the samples~\cite{grigo, kosh05B, matu01}, namely pancake vortices
with the field component perpendicular to the $ab$-plane, and
Josephson vortices with the component parallel to the $ab$-plane,
as shown in the inset of Fig.~\ref{Bz}. By rotating the field
 from in-plane to out-of-plane, the vortices change
from Josephson vortex to pancake vortex type and vice versa.

\begin {figure}[!tbp]   %fig 2
\begin{center}
\includegraphics*[bb=40 448 444 747, width=7.5cm, clip]{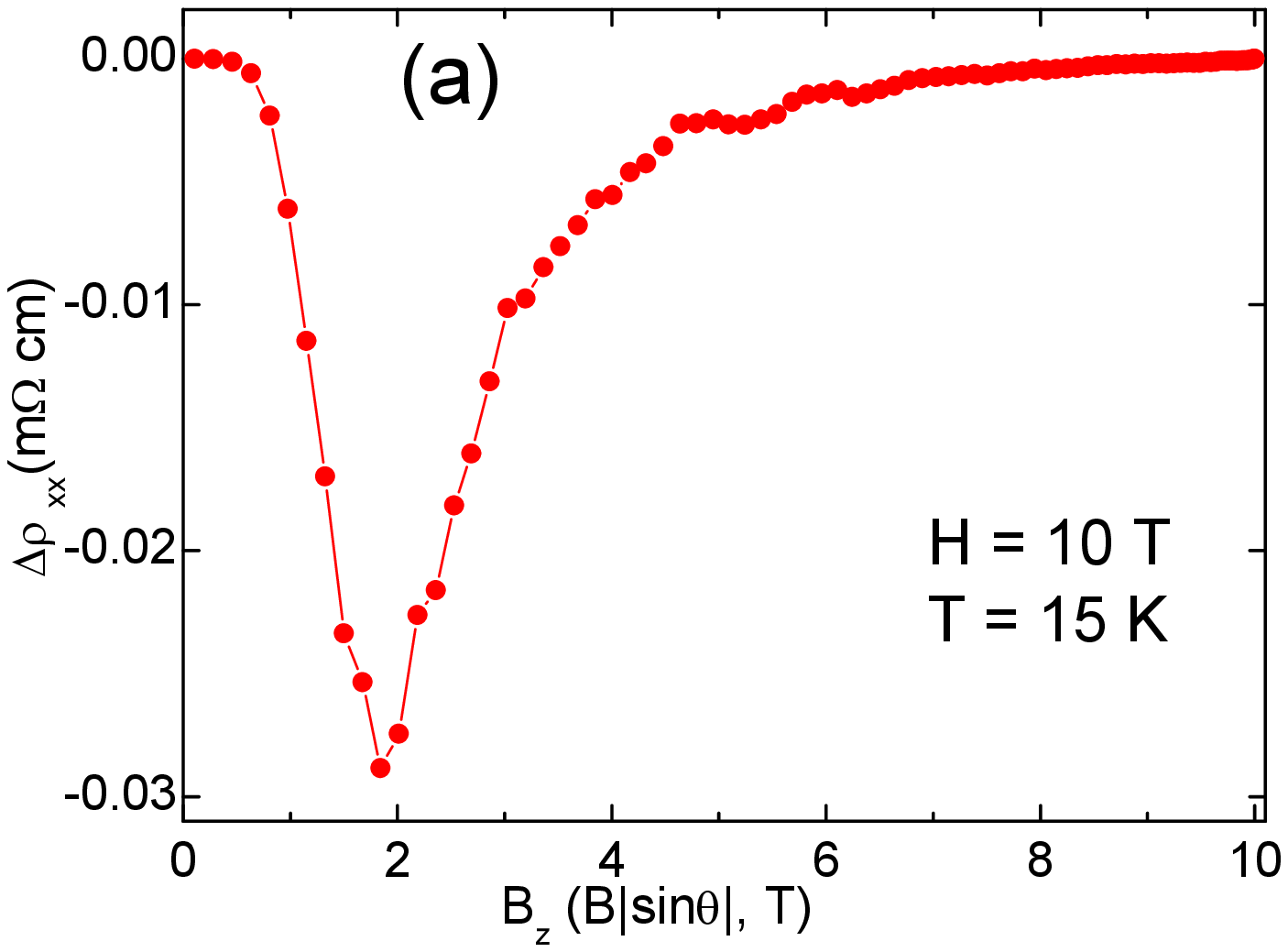}
\includegraphics*[bb=40 448 444 747, width=7.5cm, clip]{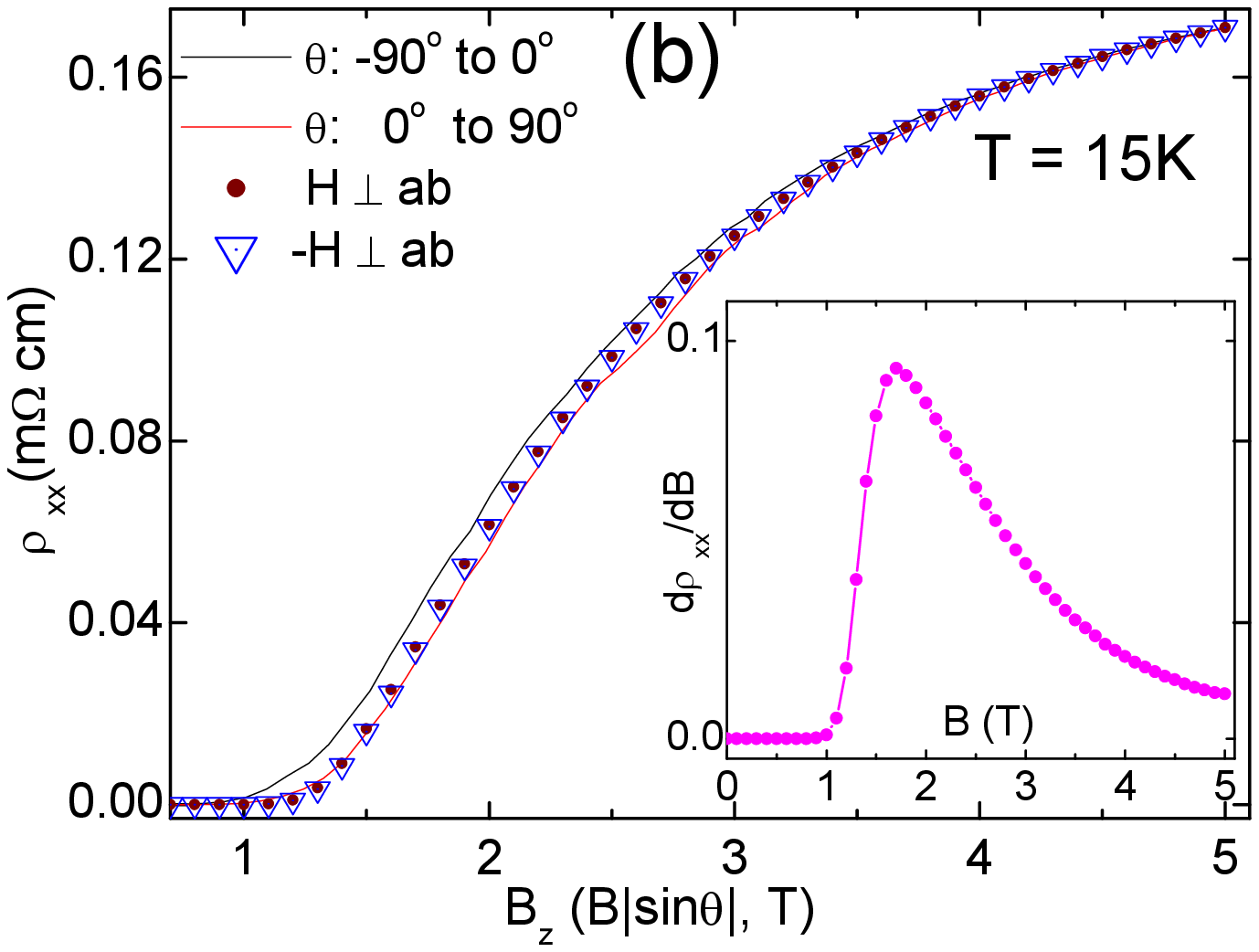}
\end{center}
\caption{(Color online) (a) The resistivity difference between the
different field orientation, i.e., $\triangle \rho_{xx}= \rho_{xx,
+ \theta}-\rho_{xx, - \theta}$, versus the field $B_z$ at H=10~T
and T=15~K. (b) The longitudinal resistivity $\rho_{xx}$ versus
the field $B_z$ in LCCO films at T=15~K by different changjing methods.
The insert in (b) shows $d \rho_{xx} / dB$ versus the magnetic
field $B$ perpendicular to the $ab$-plane. } \label{RD}
\end{figure}

In principle, the pinning effect for the in-plane vortices is
always stronger than that for out-of-plane vortices, i.e., it is
harder to drive the Josephson vortices out of the $ab$-plane.
Therefore, a hysteresis in the depinning field may
appear~\cite{amir}. Based on this consideration it is clear that
changing the field orientation from $\theta = - 90^{\circ}$ to
$\theta = 90^{\circ }$, we should be able to observe the
difference between $\rho _{xx, + \theta }$ and $\rho _{xx,
-\theta}$ due to the different evolution processes of the
vortices. Here, $\rho _{xx, +\theta}$ and $\rho _{xx, -\theta }$
correspond to the resistivity values with $\theta$ varying from
$0^{\circ}$ to $90^{\circ}$ and from $-90^{\circ}$ to $0^{\circ}$,
respectively. The experimental data seem to support this idea. In
Fig.~\ref{RD} (a), the difference between $\rho _{xx,+\theta }$
and $\rho _{xx,-\theta }$, i.e., $\triangle \rho _{xx} = \rho
_{xx, +\theta }-\rho _{xx, -\theta }$, is plotted as a function of
$B |sin \theta|$ at 15 K with fixed magnetic field strength $H =
10$~T. The maximum $| \triangle \rho_{xx} |$ can reach up to 0.03
$m\Omega \cdot cm$ at $B |sin \theta| \sim 1.8$ T ($\theta \sim
10^{\circ}$), which is about $18 \%$ of the normal-state
resistivity. With the increase of the field component along the
$c$-axis, the sample is gradually driven into the normal state and
$\triangle \rho_{xx}$ approaches zero. This demonstrates that the
in-plane field component plays an important role in the mixed
state.

In Fig.~\ref{RD}~(b), the curves of $\rho_{xx}$ versus $B | sin
\theta |$ are shown for different modes, i.e., either rotating the
orientation or changing the strength of the field. It is clear
that for the same $B_z$, the critical vortex depinning field for
mode A with $H \perp ab$ is higher than that for mode B by
rotating field, where there is additional field component parallel
to the $ab$-plane. The most important feature here is that the
critical vortex depinning field for $\rho _{xx, +\theta}$ from
$0^{\circ}$ to $90^{\circ}$ is obviously larger than that for
$\rho _{xx, -\theta}$ from $-90^{\circ}$ to $0^{\circ }$.

%-------------------------------------------------------------------------
%--------------------------------------------------------------------------

It is important to mention that the $\rho_{xx}$ curve obtained via
mode A for $H \perp ab$ overlaps with the one corresponding to $
-H \perp ab$, which suggests that the $B_z$ component solely
should not result in the hysteresis. Meanwhile, if we only apply
the in-plane field, Hall signal is not expected. So the
interaction between Josephson vortices and pancake vortices is
essential for the sign reversal of $\rho _{xy}$ in the mixed state
via rotation of the field. We would like to point out that the
field corresponding to the maximum $| \triangle \rho_{xx} |$ is
just the field where the vortices move from the plastic to the
elastic mode~\cite{Yeh} as seen from the $d\rho_{xx} / dB$($B$)
curve in the inset of Fig.~\ref{RD}(b).

\begin {figure}[!tbp]        %fig 3
\par
\begin{center}
\includegraphics*[bb=53 281 524 747, width=7.8cm, clip]{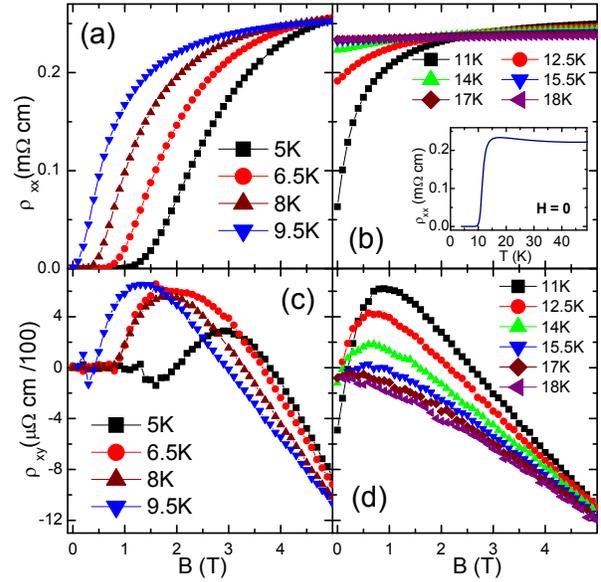}
\end{center}
\caption{(Color online) Longitudinal resistivity $\rho_{xx}$
(a)-(b) and the Hall resistivity $\rho_{xy}$ (c)-(d) versus the
magnetic field perpendicular to the $ab$-plane of LCCO:Co thin films at
different temperatures. Inset of (b): $\rho_{xx}$ versus $T$ at
zero field.} \label{RHCo}
\end {figure}

An alternative method to modify the pinning properties of the
sample consists of doping the LCCO compound with Co atoms. In this
case the Co substitution in place of Cu leads to a substantial
reduction of the critical temperature with the consequent
decrement of the critical current. Fig.~\ref{RHCo} shows
isothermal curves $\rho _{xx}(B)$ and $\rho _{xy}(B)$ for LCCO:Co
obtained with the applied magnetic field perpendicular to the
$ab$-plane. Fig.~\ref{RHCo}(a) and (c) show both $\rho_{xx}(B)$
and corresponding $\rho _{xy}(B)$ curves at temperatures below
$T_{c0}$, i.e., the critical superconducting transition
temperature at zero field. In this temperature range, and for low
enough fields vortices remain pinned thus leading to $\rho
_{xx}(B)$ = $\rho _{xy}(B)$ = 0. As $B$ increases, $\rho _{xx}(B)$
follows a smooth increase towards the normal state resistivity
value whereas $\rho _{xy}(B)$ exhibits a more complex dependence
with one ($T>6$K) or two ($T=5$K) sign reversals.  It is important
to mention that we do not observe double sign reversal in the
mixed state of LCCO~\cite{kui07B}. For temperatures higher than
$T_{c0}$ [see Fig.~\ref{RHCo}(b) and (d)], $\rho_{xy}(B)$ exhibits
a finite value at $B=0$, as shown in Fig.~\ref{RHCo}(d). This is
not surprising since the interaction between ferromagnetic Co
atoms leads to a ferromagnetic state which in turn produces a
finite internal magnetic field, even though the external magnetic
field is zero\cite{kui06B, kui07C}.

%-------------------------------------------------------------------
%--------------------------------------------------------------------
As we have demonstrated in a previous report~\cite{kui06B}, the
saturation magnetic field for LCCO:Co thin films is $\sim 0.1$~T.
This saturation field is much smaller than the values ($\sim 1$ to
4~T) at which the sign reversal occurs, which suggests that there is
no correlation between the intrinsic ferromagnetic state and the
double sign reversal. Nevertheless, the Co ions should have
preferentially the same magnetization polarity as the vortices in
this field region. The doping by magnetic ions can significantly
weaken the pinning strength in the LCCO thin films, which results in
the clear transition of the sign reversal of the Hall resistivity
$\rho_{xy}$ from once to twice as we have discussed above.

\begin {figure}[!tbp] %fig 4
\begin{center}
\includegraphics*[bb=44 450 440 747, width=7.5cm, clip]{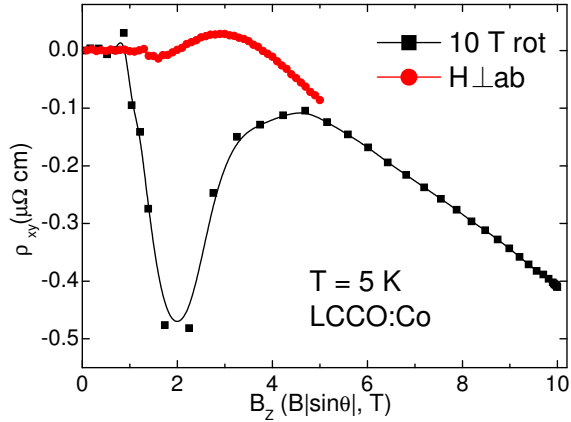}
\end{center}
\caption{(Color online) Hall resistivity $\rho_{xy}$ versus $B_z$
($B | \sin \theta |$) in LCCO:Co thin films at 5~K via different
methods, i.e., by rotating the field orientation or changing the
field strength but keeping $B \perp ab$-plane.} \label{BzCo}
\end{figure}

In Fig.~\ref{BzCo}, we show the Hall resistivity $\rho_{xy}$
versus $B_z (B | sin \theta|)$ in LCCO:Co thin films with
different modes of varying the magnetic field.
It can be seen that the minimum and maximum value with rotating
field are shifted to $\sim $ 1.9 and 4.5~T, which are larger than
those at 1.5 and 3~T by mode A with ${\bf B} \perp ab$-plane.
Besides that, the value of $\rho_{xy} (B_z)$ with rotating field
is much larger than that by mode A with ${\bf B} \perp ab$-plane,
which can be attributed to not only the coupling between the
pancake vortices and the in-plane Josephson vortices, but also to
the presence of the remanence magnetization in LCCO:Co samples
during the rotation of the magnetic field.
%
%--------------------------------------------------------------------------
%-------------------------------------------------------------------------

\section {Conclusion}

In summary, we have successfully introduced two different methods
to tune the pinning strength in the samples and we do find
significant changes in the Hall sign reversals from once to twice,
which coincides with the change of the pinning environment from
strong to weak. Our experimental results are in agreement with the
prediction of the theoretical and numerical works~\cite{wdt,
zhu96}.
The double sign reversal in the Hall resistivity has been revealed
in the electron-doped LCCO thin films via two different modes in
the present investigation. In the case of substituting Co for Cu
in LCCO thin films, the appearance of the double sign reversal of
the Hall resistivity is consistent with a weaker pinning scenario.
While, by tilting the magnetic field in the LCCO thin films, the
double sign reversal of $\rho_{xy}(B)$ is caused by the coupling
between the pancake vortices and in-plane Josephson vortices.

\section {Acknowledgments}

We acknowledge the support from the MOST 973
projects No. 2011CBA00110 and 2009CB930803, and the National
Natural Science Foundation of China and the bilateral
China/Flanders Project. This work was also supported by Methusalem
Funding by the Flemish government, FWO-Vlaanderen, and the Belgian
Inter-University Attraction Poles IAP Programmes. A.V.S. is
grateful for the support from the FWO-Vlaanderen.

\end{document}